# Primary Breadth-First Development (PBFD): An Approach to Full Stack Software Development

Dong Liu
IBM Consulting
Washington DC, USA
dliu@us.ibm.com

*Abstract*— Full stack software applications are often simplified to basic CRUD operations, which can overlook the intricate principles of computer science necessary for addressing complex development challenges. Current methodologies frequently fall short in efficiency when managing these complexities. This paper presents an innovative approach that leverages foundational computer science principles, specifically using Directed Acyclic Graphs (DAGs), to model sophisticated business problems. We introduce Breadth-First Development (BFD), Depth-First Development (DFD), Cyclic Directed Development (CDD), Directed Acyclic Development (DAD), Primary BFD (PBFD), and Primary DFD (PDFD), to enhance application development. By employing bitmaps, this approach eliminates junction tables, resulting in more compact and efficient data processing within relational databases. Rigorous testing and over eight years of production deployment for tens of thousands of users have yielded remarkable results: zero bugs, development speed improvements of up to twenty times, performance gains of seven to eight times, and storage requirements reduced to one-eleventh compared to traditional methods.

I. INTRODUCTION

Full stack software development typically encompasses both front-end and back-end components. By leveraging full stack development, businesses can create robust, scalable, and user-friendly applications that meet diverse needs, from e-commerce and social media to healthcare and government.

Many full stack applications are simple CRUD applications powered by relational databases. Since their invention in the 1970s, relational databases have remained the most widely used type of database [1][2][3] due to their structured data integrity and accuracy. Consequently, many companies specifically request the use of RDBMS. However, relational databases present challenges when dealing with complex hierarchical data, especially in full stack applications developed under tight deadlines and limited budgets.

Consider the development of a website to log visitors to various global locations. Users follow these steps from the UI:

1. Users select one or more of the seven continents: Africa, Antarctica, Asia, Australia, Europe, North America, and South America.

2. Users then select the corresponding countries within each chosen continent, except for Antarctica, where they select research stations instead.

3. The subsequent steps vary by country. For example, in the United States, users would choose: State, County, and City. In China, the steps would be Province, City, and County.

Fig. 1 illustrates a simplified UI structure of this system, forming an n-ary tree. Each level corresponds to a step.

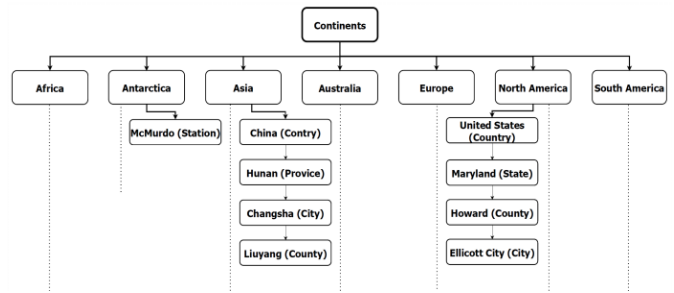

Fig. 1. Business Model for Logging Visited Places

In Fig. 1, three complete branches are shown, but there could be thousands. Not all branches share the same structure; for example, Antarctica has only one level beneath it, while China and the United States have three additional levels with different sequences between city and county.

Various methods, including Nested Sets [4], Adjacency Lists [5], Closure Tables [6], and Materialized Paths [7], can model these data structures in a relational database. One approach is illustrated in Fig. 2, which uses the Adjacency Lists model in the Location table.

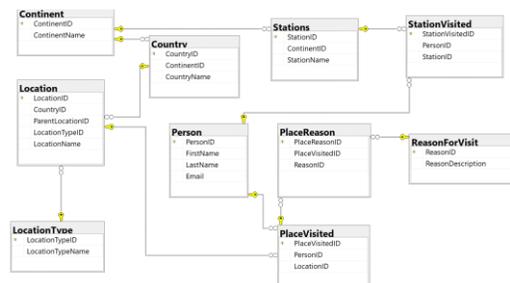

Fig. 2. Traditional database schema

When normalizing data, several factors must be considered. Understanding the business context of the data is essential for decomposing and reorganizing it into a suitable schema for storage. Often, the data must be denormalized for front-end display while remaining normalized for storage. This design and development process is both time-consuming and complex.

In a client project, we were tasked with developing web pages more complex than the example above, all within a three-week timeline. Completing these pages—covering front-end, back-end, and database components—within this timeframe while maintaining quality proved nearly impossible for a single developer using traditional full stack methods. A novel approach was necessary.

## II. Definitions

Cycle Graph, Directed Acyclic Graph (DAG), Depth First Search (DFS) and Breadth First Search (BFS) are fundamental computer science concepts widely used in data structures and algorithms [8][9][10][11]. However, these graph techniques are rarely applied directly in full stack software development to describe software design and development steps.

To address this, we introduce analogous concepts for the development process, integrating graph theory principles into a structured software development approach.

- Depth-First Development (DFD) focuses on completing all levels of a specific path before moving to the next. Development proceeds deeply along each path, ensuring it is fully developed before advancing.

- Breadth-First Development (BFD) addresses all neighboring vertices in the current level before progressing. It emphasizes handling similar features while advancing level by level, typically starting from the source vertices and moving downward, ensuring balanced and cohesive system evolution.

- Cyclic Directed Development (CDD) combines iterative cycles with a directed sequence of development stages, revisiting previous vertices and incorporating feedback loops for continuous improvement.

- Primary DFD (PDFD) focuses on DFD, with occasional BFD or CDD at certain levels.

- Primary BFD (PBFD) prioritizes BFD, occasionally applying DFD or CDD after developing vertices with similar features within the same level.

- Directed Acyclic Development (DAD) applies DAG principles to structure and manage tasks. Each task, feature, or component is represented as a vertex in the DAD, with directed edges showing dependencies and execution order. DAD created the model in Fig. 1.

## III. Traditional Implementation

Modern Agile methodologies promote an iterative and incremental development approach [12]. Initially, they align with DFD principles, focusing on delivering a quick, functional end-to-end feature. However, they sometimes incorporate a level-by-level sequence similar to BFD, blending into what we refer to as PDFD. Over time, these iterations converge into CDD.

For example, in place-visited development (Fig. 1), we might start by developing a continents page, followed by a North America countries page, a United States states page, a Maryland counties page, and finally a Howard cities page. Afterward, we continue to other Maryland counties' cities pages. This approach aligns with PDFD principles, where each page may not be fully developed in the first pass. The primary goal is to complete a path from the source to the leaves before moving on to others. During Scrum, the process may shift to CDD, revisiting previously processed vertices in iterative cycles.

Revisiting and extending paths often requires returning to earlier stages, adding or modifying code in the same files, modules, classes, and functions. This leads to rewriting and retesting processed vertices, creating redundant work. As a result, the development time complexity can approach $O(V^2)$, where V is the number of vertices.

## IV. Our Implementation

We employ a PBFD approach, leveraging level-wise bitmaps for data compression. This results in a unique database design, with views added to PBFD to interpret the bitmaps and their relationships with other data. When bitmaps are unsuitable, other data structures, such as hash tables, may be utilized.

### A. Bit Representations of Neighboring Vertices

We use bitmaps to represent neighboring vertices when they can be grouped into 32-bit or 64-bit integers. For example, in a 32-bit bitmap for continents, 'Africa' is represented as '0000 0000 0000 0001'. Shifting one bit left represents the next continent: 'Antarctica' becomes '0000 0000 0000 0010', and 'Asia' is '0000 0000 0000 0100'. These values are represented in the IdContinent column of a lookup table (Table I).

TABLE I. Continent Lookup Table for PBFD

| IdContinent | Name |
|---|---|
| 1 | Africa |
| 2 | Antarctica |
| 4 | Asia |
| 8 | Australia |
| 16 | Europe |
| 32 | North America |
| 64 | South America |

The selected values for all continents are combined into a single value using a bitwise OR operation. For example, if all seven continents are selected, the resulting bitmap is "0000 0000 0111 1111," corresponding to a value of 127. If only "Asia" and "North America" are selected, the bitmap is "0000 0000 0010 0100," resulting in a value of 36.

### B. PBFD

Neighboring vertices often share properties (e.g., visa requirements at the country level). We process vertices with similar features at each level before advancing, allowing for efficient development with shared and sealed code. This approach makes data handling more compact and manageable. For example, we consolidate all continents into components using bitmap representations and process child vertices similarly, creating an Africa bitmap for African countries, an Asia bitmap for Asian countries, and so on, continuing in a breadth-first manner through subsequent levels.

However, we do not strictly adhere to the BFD approach. When necessary—such as to complete a critical feature or demonstrate an end-to-end path to business stakeholders—we adopt DFD to finalize relevant subset features under specific vertices, like the United States, before returning to BFD for other countries. During Scrum, we incorporate CDD to address evolving business needs. Thus, our approach is a flexible PBFD method that adapts as required.

### C. Database Layer

Fig. 3 illustrates the database model. The Continent table stores data presented in Table I, which corresponds to the checkboxes or dropdown select options in the front end. The VisitedLocation table captures user selections from the UI, storing them as bitmaps. For example, IdContinents in the VisitedLocation table could store the bitmap value of 127 for all continents or 36 for only Asia and North America.

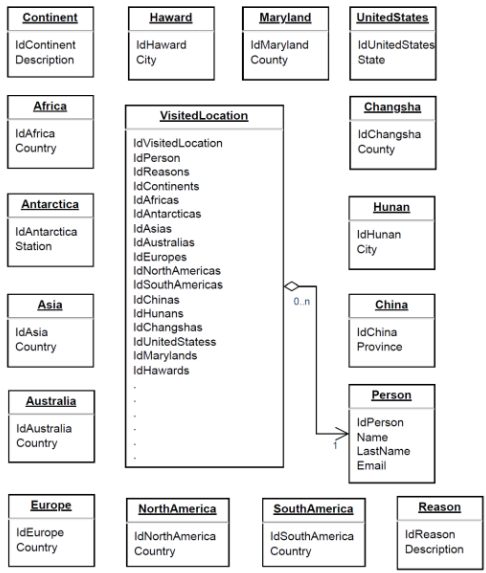

Fig. 3. Database schema for the PBFD approach.

The VisitedLocation table follows a breadth-first sequence scan from the UI through the business layer, storing data in a compact and sequential format starting with IdContinents. This approach minimizes the need for complex data mappings and conversions between the UI, business, and database layers.

The columns expand like a flattened n-ary tree, with parent columns storing bitmap data that represent edges connecting to child columns. For instance, IdChinas holds a bitmap of visited provinces in China, while IdUnitedStates contains a bitmap for visited states in the U.S. For simplicity, Fig. 3 shows only one child column for each country: IdHunans and IdMarylands.

IdReasons contains the same data as the PlaceReason many-to-many junction table in traditional database designs (see Fig. 2). PBFD simplifies the traditional relational database by converting tables into columns and columns into cells, effectively reducing database dimensions and complexity.

PBFD continues to leverage relational database normalization and other principles when it is necessary and efficient. In Fig. 2, the Person table has a referential relationship with VisitedLocation, but other lookup tables do not. These relationships are managed by an n-ary tree within the business layer. PBFD integrates the database layer with the business layer to handle these relationships.

### D. Business Layer

The business layer loads and caches lookup tables in memory. Our applications are developed using ASP.NET MVC, with the checkboxes' view model defined in C# as follows:

```
1. class CheckBoxListItem{
2.     int ID;
3.     string Description;
4.     bool    IsChecked;    //
   Indicates           whether  the
   checkbox is checked (default is
   false).
5.     string Class // Name of the
   CSS class used for UI.
6. }
```

On the Continent checkboxes page, the instances of the above class will be populated with data where the ID is mapped to the IdContinent column in the Continent table, and the IsChecked property is mapped to the IdContinents column in the VisitedLocation table. The controller sends a list of these view model objects to the front end, where their content is displayed.

When the user updates the checkboxes, the modified CheckBoxListItem objects are sent to the backend, where they are converted into a bitmap using the following code:

```
1. int
   CalculateIDs(IList<CheckBoxListItem
   > listObj){
2.     int ids = 0;
3.     foreach (int obj in listObj)
4.     {
5.         if (obj.IsChecked)
6.             ids |= obj.ID;
7.     }
8.     return ids;
9. }
```

For the continent page, ids are stored as IdContinents in the VisitedLocation table. Both the frontend and backend employ a breadth-first approach to data processing. The business layer utilizes data structures such as bitmaps, n-ary trees, and hash tables to manage edges, vertices, and parent-child relationships, mapping them to the database level by level. The reading process of the application simply reverses the writing process.

*E. Database Report*

Unlike traditional designs that rely on junction tables (Fig. 2), our approach enhances data persistence by reducing their usage. However, this efficiency limits the ability to generate traditional relational database reports, which often rely on the primary and foreign key referential of base and lookup tables. Consequently, the data hierarchy cannot be directly constructed from the database schema (Fig. 3) alone. Following the DAG business model in Fig. 1 and using the breadth-first process sequence, we utilize database views to reconstruct data in a Materialized Path-like manner from the schema. For example, if Ellicott City, Liuyang, and McMurdo in Fig. 1 are visited, the paths from the source to these leaves are combined in the views, with each column representing a level of the n-ary tree. Reporters can query these views instead of the underlying tables.

These views are created lazily, meaning they are only generated when required for reporting, which may be months or years after the system goes live.

## V. RESULTS

*A. Superior Development Speed*

Traditional full-stack software development typically has time complexity ranging from $O(V)$ to $O(V^2)$. In the best-case scenario, similar to Depth-First Search (DFS), each vertex is visited once, resulting in $O(V)$. However, in the worst case, vertices may be revisited multiple times, especially when a source vertex is revisited with each addition to the structure, pushing the complexity toward $O(V^2)$. Reiteration, commonly seen in Agile methodology, increases the likelihood of reaching $O(V^2)$.

After business requirements are translated into an n-ary tree by the DAD (Fig. 1), PBFD processes this structure using a breadth-first strategy across the UI, business, and database layers. This results in a more efficient development process with $O(V)$ time complexity.

With PBFD, we maintain a breadth-first traversal across the UI, business layer, and database stages, resulting in enhanced efficiency with a time complexity of $O(V)$ in most cases. In 2016, we implemented the PBFD approach for a client's complex hierarchical data on web pages, completing the feature in approximately one month (June 2, 2016 – July 5, 2016) with a single developer working at a time.

From July 21, 2021, to April 27, 2022, we converted the PBFD database design for the same feature to a traditional relational database design and migrated the data. This conversion and SQL code development took two developers working part-time over nine months, excluding front-end and other back-end tasks. This suggests that PBFD could be at least nine times faster than traditional full-stack software development in an Agile environment.

From August 20, 2022, to June 30, 2024, we transformed this feature into a similar solution using Salesforce OmniScript. The process took about two years and involved approximately seven developers. The project was still ongoing at the time this paper was written. We estimate that PBFD development speed is more than 20 times faster than the Salesforce approach.

A scientific and accurate quantitative evaluation of development speed in the real world is challenging, as many factors can affect development speed in a competitive business environment. However, the development approach remains a critical determining factor. While we need to code all layers as in the traditional approach, our method achieves exceptional speed likely due to several key factors:

1. Use of Graph Theory to Model Business Scenarios (Fig. 1): This approach turns complex business processes into a structured, clear, and manageable development blueprint. Developers simply follow the blueprint step-by-step. Without this, developers can become overwhelmed by development requirements and lose sight of both the big picture and the fine details, leading to exponentially increased development time as complexity grows. This was evident in our Salesforce team's experience. Although Salesforce's OmniScript approach requires minimal coding, the components needed to handle the complexity of highly hierarchical data presented additional challenges.

2. PBFD's Emphasis on Sequential Development of Similar Content: PBFD allows developers to maintain a consistent development context, reducing the need for frequent context-switching. Developers can focus on the content and produce fewer bugs in each application layer, class, module, and file. Once a step is completed, it does not require repeated testing.

3. Specific Techniques Contributing to Fast Development: Our use of bitmap data compaction simplifies data handling significantly. Additionally, the database design is much simpler, eliminating the need for complex table relationships. When business entity relationships are intricate, significant time is spent designing relational databases. PBFD reduces this complexity.

*B. Excellent Application Performance*

Both traditional full stack software development and PBFD have a time complexity of $O(V)$ in application execution. However, PBFD eliminates many-to-many and one-to-many database joins, which increases the likelihood of sequential data writing to the database, significantly improving efficiency.

Over eight years of client log data, PBFD delivered application speeds seven to eight times faster than traditional methods, based on mean, median, and 95th percentile metrics. Additionally, PBFD produced no reported bugs, deadlocks, or performance issues, unlike traditionally developed pages.

*C. Smaller Database Storage*

In traditional full stack development, the space complexity for database writes is $O(V)$, as each vertex is written once or a constant number of times. PBFD also maintains $O(V)$ but uses 64-bit or 32-bit bitmaps to represent data more efficiently, reducing storage by a factor of 64 or 32 for binary data.

In our client implementation, we found that the database size using the traditional method was approximately eleven times larger than the compressed size achieved with PBFD.

### D. Simpler Design and Development

PBFD follows a Breadth-First approach, which simplifies the design by progressing systematically from the front end to the back end. This method is intuitive, mirroring the process of scanning an image pixel by pixel and line by line using run-length encoding. It minimizes the frequent context switches often associated with DFD or Continuous CDD. This approach minimizes the risk of overlooking business scenarios or introducing bugs, thereby speeding up development. A developer new to PBFD on the client project needed only thirty minutes of training to complete a new feature within a week.

PBFD also eliminates the need for certain junction tables commonly used in traditional relational database designs, thereby reducing complexity in both design and programming.

### E. Better Separation of Concerns

Writing to a database is more complex than reading from it, as it involves transaction management, index updates, data validation, logging, concurrency control, disk I/O, replication, and backups. Therefore, separating the processes of data collection (write operations) from report generation (read operations) can help avoid the issues that arise when using the same schema for both in traditional relational databases.

PBFD prioritizes efficient data collection, utilizing computer-friendly bit operations without being concerned with the complexity of report generation. When needed, PBFD provides views that are human-friendly, simplifying the relational database's complexity. PBFD views were chosen for the client's data migration due to their simplicity, reliability, and efficiency, demonstrating superiority over traditional methods.

## VI. CONCLUSIONS

Revisiting fundamental computer science concepts and introducing new ones is crucial for tackling complex problems lacking efficient solutions. This paper addresses intricate business challenges using DAG structures and introduces novel concepts such as BFD, DFD, PBFD, PDFD, CDC, and DAD. It demonstrates PBFD's implementation through simplified examples involving hierarchical data.

The PBFD implementation has proven highly effective for our client, significantly accelerating development speed and achieving zero bugs over eight years for tens of thousands of users. It also enhances application performance compared to traditional methods and reduces storage requirements. PBFD presents a potential solution to the challenges faced in modern software development.

Additionally, the design patterns and code from PBFD are easier for developers to follow and extend, and database developers favor PBFD views over traditional table joins. The methodologies discussed can be generalized to broader software development contexts, albeit with variations in scale.

## VII. FUTURE PLANS

1. We will continue working with our client to obtain permission for publishing a full-length paper.
2. We plan to extend the use of PBFD to other appropriate applications.

## VIII. ACKNOWLEDGEMENTS

The author wishes to express gratitude to his IBM managers, Jen Kostenko, Ricardo Zavaleta Cruz, and Anton Cwu, for their support, for reviewing this paper, and for granting approval for its publication.